\documentclass[aps,prl,preprint,superscriptaddress,floatfix]{revtex4-2}

\usepackage{graphicx}
\graphicspath{ {./images/} }
\usepackage{xcolor}
\usepackage{amsmath}

\begin{document}


\title{Emergent Magnetic Phases and Piezomagnetic Effects in Mn$_x$Ni$_{1-x}$F$_2$ Thin Film Alloys}

\author{Ryan Van Haren}
\affiliation{Department of Physics, University of California Santa Cruz, Santa Cruz, California 95064}
\email{rvanhare@ucsc.edu}
\author{Nessa Hald}
\affiliation{Materials Science and Engineering Program, School for Engineering of Energy, Matter, and Transport, Arizona State University, Tempe, AZ 85287, USA}
\author{David Lederman}
\affiliation{Department of Physics, University of California Santa Cruz, Santa Cruz, California 95064}


\date{\today}

\begin{abstract}

The effect of random competing single-ion anisotropies in antiferromagnets was studied using epitaxial Mn$_x$Ni$_{1-x}$F$_2$ antiferromagnetic thin film alloys grown via molecular beam epitaxy. The crystal structure of this material is tetragonal for all values of $x$, and the Mn sites have a magnetic easy axis single-ion anisotropy while the Ni sites have an easy plane anisotropy perpendicular to the Mn easy axis. Crystallographic and magnetization measurements demonstrated that the thin film alloys were homogeneously mixed and did not phase-separate into their constituent parts. Pure MnF$_2$ thin films epitaxially grown on MgF$_2$ exhibited compressive strain along all three crystallographic axes which resulted in piezomagnetic effects. The piezomagnetism disappeared if the film was grown on a (MnNi)F$_2$ graded buffer layer. A mean-field theory fit to the transition temperature as a function of the Mn concentration $x$, which takes into account piezomagnetic effects, gave a magnetic exchange constant between Mn and Ni ions of $J_{\text{MnNi}} = 0.305 \pm 0.003$~meV. Mean-field theory calculations also predicted the existence of an oblique antiferromagnetic phase in the Mn$_x$Ni$_{1-x}$F$_2$ alloy which agreed with the experimental data. A magnetic phase diagram for Mn$_x$Ni$_{1-x}$F$_2$ thin film alloys was constructed and showed evidence for the existence of two unique magnetic phases, in addition to the ordinary antiferromagnetic and paramagnetic phases: an oblique antiferromagnetic phase, and an emergent magnetic phase proposed to be either a magnetic glassy phase or a helical phase. The phase diagram is quantitatively different from that of Fe$_x$Ni$_{1-x}$F$_2$ because of the much larger single-ion anisotropy of Fe$^{2+}$ compared to Mn$^{2+}$.

\end{abstract}


\maketitle

\section{Introduction}

NiF$_2$ and MnF$_2$ are model antiferromagnets (AFs) which share a rutile, tetragonal P4$_2$/mnm space group crystal structure with similar lattice parameters~\cite{stout_crystal_1954}, but which have different magnetic structures. NiF$_2$ is a (110) easy plane antiferromagnet with an effective Dzyaloshinskii–Moriya interaction (DMI) that causes the antiferromagnetically aligned moments to spontaneously cant in the easy plane, generating a weak ferromagnetic moment perpendicular to the N\'eel vector \cite{dzyaloshinsky_thermodynamic_1958,moriya_theory_1960,borovik-romanov_weak_1973}. The DMI is an important ingredient for the development of stable helical spin textures, such as skyrmions and hopfions, which have promising applications in spintronic technologies \cite{emori_current-driven_2013,barker_static_2016, zhang_antiferromagnetic_2016,kent_creation_2021,rybakov_magnetic_2022}. MnF$_2$ lacks a DMI, and its [001] easy axis is due primarily to dipole-dipole interactions resulting from the crystal structure of the material, making it a useful system in which to study magnons in Ising-like systems \cite{barak_magnetic_1978, wu_antiferromagnetic_2016, zhao_magnon_2006}. MnF$_2$ also has a relatively small and accessible spin-flop field ($\sim 9.3$~T), making it easier to perform steady-state measurements above the spin-flop transition \cite{jacobs_spin-flopping_1961, felcher_antiferromagnetic_1996}. 

The Mn$_x$Ni$_{1-x}$F$_2$ alloy system is interesting because of the competing single-ion magnetic anisotropies of the Ni$^{2+}$ and Mn$^{2+}$ ions (perpendicular to, and along the [001] direction, respectively). The resulting random magnetic anisotropy can lead to new magnetic phases near the critical temperature of the material, where the single-ion anisotropy term dominates the spin Hamiltonian. One such material that possesses these characteristics, Fe$_{x}$Ni$_{1-x}$F$_2$, has been studied previously and exhibited a unique magnetic phase diagram with evidence of a magnetic glassy phase due to random magnetic anisotropy \cite{perez_phase_2015}. MnF$_2$ is similar to FeF$_2$ in that it has the same rutile crystal structure and is an easy axis AF that orders along the $c$-axis, but it has a spin of 5/2 instead of 2 and importantly, it has a single-ion anisotropy energy that is nearly 10 times smaller than that of FeF$_2$ \cite{stout_crystal_1954,hutchings_spin_1970,barak_magnetic_1978}. This makes the Mn$_x$Ni$_{1-x}$F$_2$ alloy an interesting point of comparison with previous work on Fe$_{x}$Ni$_{1-x}$F$_2$ as it demonstrates how differences in the single-ion anisotropy energy affect the magnetic properties of the system. Understanding how this parameter affects the system is essential to accurately predicting ordering behavior near the transition temperature because the single-ion anisotropy energy will dominate the spin Hamiltonian near this critical point \cite{perez_phase_2015}.

This work presents a crystallographic and magnetic study of Mn$_x$Ni$_{1-x}$F$_2$ thin film alloys. We find that the Mn$_x$Ni$_{1-x}$F$_2$ thin film alloys are mixed homogeneously and do not separate into their constituent NiF$_2$ and MnF$_2$ parts. We also find that epitaxial MnF$_2$ thin films grown on MgF$_2$ are highly strained, which has the effect of lowering the AF transition temperature due to piezomagnetism~\cite{borovik-romanov_piezomagnetism_1960,baruchel_180_1980,baruchel_piezomagnetism_1988}. This epitaxial strain induced piezomagnetism is verified in a relaxed MnF$_2$ thin film grown using a (MnNi)F$_2$ graded buffer layer, from which we find that when epitaxial strain is eliminated, the transition temperature of the relaxed MnF$_2$ thin film matches the expected bulk value. Magnetization measurements of Mn$_x$Ni$_{1-x}$F$_2$ thin film alloys show that the system has a rich magnetic phase diagram, including an emergent ordered unidentified phase in a narrow temperature range near the transition temperature. Mean-field theory (MFT) equations using the true random magnetic anisotropy are presented and are compared with the experimentally derived phase diagram and exchange energies of the thin film alloys.

\section{Experimental Methods}
The Mn$_x$Ni$_{1-x}$F$_2$ alloy thin films in this study were all grown in an ultra-high vacuum molecular beam epitaxy (MBE) system (base pressure $< 10^{-8}$ Torr) by sublimation of commercially purchased NiF$_2$ and MnF$_2$ powders ($>99$\% purity) onto commercially purchased MgF$_2$ (110) substrates. Before starting the growth process, the substrate was annealed at $T = 300 ^\circ$C  in the growth chamber for a minimum of 1 hour. Reflection high-energy electron diffraction (RHEED) patterns were acquired after annealing the substrate to ensure satisfactory surface smoothness and crystallinity before deposition. A retractable crystal monitor inside the growth chamber was used to calibrate the molecular flux of the NiF$_2$ and MnF$_2$ beams and to set the desired stoichiometry of each sample. All Mn$_x$Ni$_{1-x}$F$_2$ thin films, including the $x=0$ (pure NiF$_2$) and the $x = 1.0$ (pure MnF$_2$) films, were grown to a thickness of 30~nm after growing an epitaxial 1 nm thick NiF$_2$ buffer layer between the substrate and the alloy film, in order to reduce lattice mismatch and create a higher-quality film~\cite{perez_phase_2015}. MnF$_2$ films with reduced strain were grown using a 20~nm thick (MnNi)F$_2$ graded buffer layer, where Mn$_x$Ni$_{1-x}$F$_2$ was first deposited with $ x = 0$ (pure NiF$_2$) and then the MnF$_2$ flux was slowly increased while the NiF$_2$ flux was gradually decreased simultaneously until the top of the film was $ x = 1$ (pure MnF$_2$), at which point a 30~nm MnF$_2$ film was grown. RHEED patterns of all films were then acquired before removing the films from the vacuum system.

X-ray diffraction (XRD) measurements of the thin films were performed using Cu K$_\alpha$ radiation from a Rigaku SmartLab thin film x-ray diffractometer. The value of the (110) lattice parameter out of the plane of the sample was calculated from the XRD peak positions of the peaks according to Bragg's law, $2d_{hkl}\sin (\theta_{hkl})=\lambda$, where $d_{hkl}$ is the lattice constant corresponding to planes defined by the Miller indices $(hkl)$, $\theta_{hkl}$ is the measured Bragg diffraction angle corresponding to the $(hkl)$ plane, and $\lambda=0.15406$~nm is the x-ray wavelength used. 

Magnetic properties of the films were studied using a Quantum Design MPMS XL superconducting quantum interference device (SQUID) magnetometer by measuring magnetic moment as a function of temperature. The transition temperature of each sample was determined by fitting magnetic moment as a function of temperature near the critical point to a distribution of sharp transition temperatures due to disorder or other factors given by  
\begin{equation}
    m(T) = \frac{C}{\sigma_c \sqrt{2\pi}}\int_{T}^{\infty}(1-T/T_c')^{\beta}e^{-(T_c - T_c')^2/2\sigma_c^2}dT_c',
\end{equation}
where $C$ is a scaling parameter, $\sigma_c$ is the rounded width of the transition, $T_c$ is the average critical temperature of the sample, $\beta$ is the critical exponent, and $T_c^\prime$ is a dummy variable in the integral for the transition temperature distribution \cite{perez_phase_2015}. 

\section{Results}

\subsection{Thin Film Crystallography}

\begin{figure*}[h]
    \centering
    \includegraphics{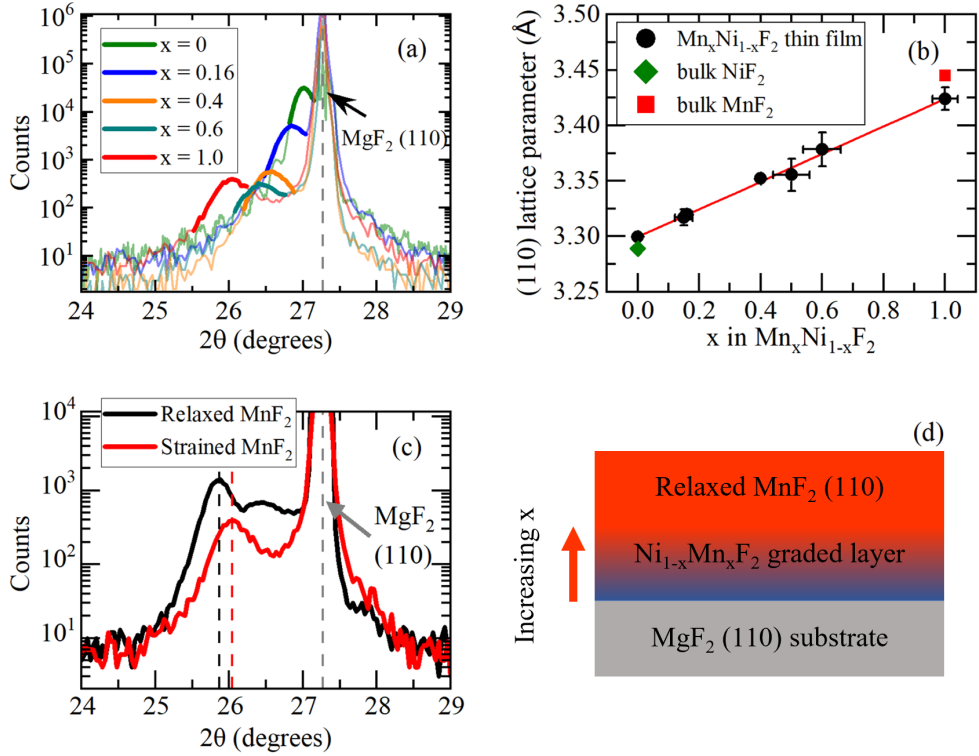}
    \caption{ \label{fig:XRD} (a) XRD pattern near the (110) peak of the MgF$_2$ substrate and Mn$_x$Ni$_{1-x}$F$_2$ thin films. (b) Calculated (110) lattice parameter as a function of Mn stoichiometry $x$. Solid red line is a linear fit to the thin film data. Bulk (110) lattice parameters of NiF$_2$ and MnF$_2$ are represented as the green diamond and red square, respectively. (c) XRD pattern near the (110) peak of strained and relaxed MnF$_2$ thin films. (d) Diagram of the relaxed MnF$_2$ thin film sample using a (MnNi)F$_2$ graded buffer layer.} 
\end{figure*}

XRD measurements of the Mn$_x$Ni$_{1-x}$F$_2$ thin films showed that the films grew in the [110] crystal orientation without any evidence of additional peaks that would indicate phase separation of the alloys into NiF$_2$ and MnF$_2$ domains. Instead, the position of the (110) peak shifted to smaller angles with increasing $x$ as shown in Fig. \ref{fig:XRD}(a). This behavior is consistent with a smoothly mixed Mn$_x$Ni$_{1-x}$F$_2$ crystal alloy, where the (110) lattice parameter corresponds to the stoichiometric average of the constituent MnF$_2$ and NiF$_2$ constituents. The (110) out-of-plane lattice parameters measured for the Mn$_x$Ni$_{1-x}$F$_2$ thin films are plotted in Fig.~\ref{fig:XRD}(b), along with the the expected (110) lattice parameters of bulk MnF$_2$ and NiF$_2$ \cite{baur_rutile-type_1976, stout_crystal_1954}. Full XRD scans from $10^\circ < 2\theta < 80^\circ$ showing single phase (110) orientation thin films are provided in the supplemental material \cite{VanHaren2023}. The calculated lattice parameter values are consistent with the claim  that there is no phase separation in the Mn$_x$Ni$_{1-x}$F$_2$ thin film alloys, as the calculated values fit nicely along a linear trend line between the $x = 0$ and $x = 1$ thin films, shown as the solid red line in Fig.~\ref{fig:XRD}(b). 

Note that the lattice parameters of the pure NiF$_2$ and MnF$_2$ endpoint samples in Fig.~\ref{fig:XRD}(b), are different from the lattice parameters of their respective bulk values. This lattice strain in the thin film samples is due to epitaxial growth on the MgF$_2$ substrate, which has a smaller unit cell than either NiF$_2$ or MnF$_2$ \cite{VanHaren2023}. Our XRD measurements indicate that there is tensile strain along the [110] growth direction in NiF$_2$ (110) thin films grown on MgF$_2$ (110), in agreement with previous studies~\cite{shi_exchange_2004}. The NiF$_2$ film experiences in plane compressive strain along the $c$ axis due to epitaxial growth on the smaller MgF$_2$ substrate, while it expands slightly along the [110] direction to accommodate this compression. 

Something unusual happens in the case of MnF$_2$ grown on MgF$_2$, as XRD measurements indicate that the crystal compresses along the [110] direction, contrary to the behavior observed in NiF$_2$. Careful x-ray measurements of the out-of-plane diffraction peaks [the (110) peak] and peaks with  in-plane components of the scattering vector [the (111) and (211) peaks] allowed us to calculate all three unit cell axes \cite{VanHaren2023}. These values are given in Table~\ref{tab:MnF2}. Our results indicate that the MnF$_2$ thin film is compressed along all three crystallographic directions due to epitaxial growth on the MgF$_2$ substrate. This is unusual as the expected behavior from crystals under strain is that the lattice will expand along some axes to compensate for compression along others in order to maintain the same unit cell volume. Our measurements indicate that the unit cell volume actually decreases by a small amount due to compressive strain along all axes. This behavior is possibly explained by the (110) crystal orientation epitaxial growth. The (110) face of the crystal has both the [001] and the [1$\bar{1}$0] crystallographic directions lying in plane with the MgF$_2$ (110) substrate. The epitaxial growth could result in both the [001] and [1$\bar{1}$0] axes feeling compressive strain at the interface and thus result in a MnF$_2$ thin film crystal with a reduced unit cell volume. While these XRD measurements were performed at room temperature well above the magnetic transition, it is reasonable to assume that the difference in strain between the two MnF$_2$ thin films remains even at low temperature. Future work could investigate how the crystal structure changes as a function of temperature, particularly near the N\'eel temperature, to test this assumption. How this strain affects the magnetization of the MnF$_2$ film will be discussed below.

\begin{table}[h]
    \caption{\label{tab:MnF2} Lattice parameters of relaxed MnF$_2$ from reference \cite{stout_crystal_1954} and strained thin film MnF$_2$ grown for this study. Lattice units are in \AA.}

\begin{ruledtabular}
\begin{tabular}{ccccc}
    Material & $a$ & $b$ & $c$ \\
\hline
   Bulk MnF$_2$ & 4.873 & 4.873 & 3.310 \\

    Thin film MnF$_2$ & $4.852 \pm 0.007$ & $4.848 \pm 0.007$ & $3.291 \pm 0.004$ \\

\end{tabular}
\end{ruledtabular}
\end{table}

In order to test if the observed strain was due to epitaxial growth on the smaller unit cell of MgF$_2$ and how this affects the magnetism of the film, comparison with a relaxed MnF$_2$ thin film is necessary. The strain observed in MnF$_2$ thin films grown on MgF$_2$ can be eliminated by the use of a (MnNi)F$_2$ graded layer as a buffer between the substrate and the MnF$_2$ thin film, as shown in Fig.~\ref{fig:XRD}(d) and described in the Methods section above. By gradually increasing the Mn stoichiometry $x$ in the buffer layer with increasing thickness, the lattice parameters of the buffer layer slowly increased, ultimately resulting in a relaxed MnF$_2$ thin film with improved crystallinity, as shown by XRD measurements in Fig.~\ref{fig:XRD}(c). It is unlikely that there is a sharp boundary between the (MnNi)F$_2$ graded layer and the pure MnF$_2$ layer, because the growth of the (MnNi)F$_2$ graded layer growth is designed in such a way that the layer smoothly transitions from NiF$_2$ to MnF$_2$ as it gradually changes the lattice parameter of the film to reduce strain between the substrate and the MnF$_2$ film, although further measurements such as transmission electron microscopy or x-ray photoelectron spectroscopy would be needed to verify the structure. A comparison of the magnetic behavior of the strained MnF$_2$ film with the relaxed film will be presented in the next section.

\begin{figure*}[h]
    \centering
    \includegraphics{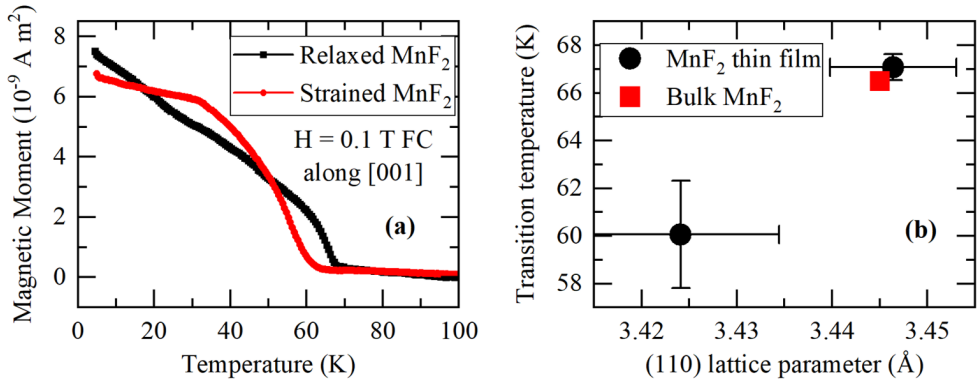}
    \caption{\label{fig:MnF2Piezo} (a) Magnetic moment measurements of strained and relaxed MnF$_2$ thin films field-cooled (FC) in a $\mu_0 H=0.1$~T external magnetic field applied along the $c$ axis.(b) magnetic transition temperature of MnF$_2$ thin films, relaxed and strained, and bulk MnF$_2$ as a function of the (110) lattice parameter. }
    
\end{figure*}

\subsection{Magnetization}
Field-cooled (FC) Magnetic moment measurements of the strained and relaxed MnF$_2$ thin films along the $c$-axis are shown in Fig.~\ref{fig:MnF2Piezo}(a), revealing a shift in the transition temperature between these two films. FC measurements are performed by warming the sample above the N\'eel temperature to $T = 100$~K then setting the external magnetic field to $\mu_0H = 0.1$~T and measuring the moment as the sample is cooled. Although it may be expected that MnF$_2$ would have no net magnetic moment along $c$-axis below the N\'eel temperature because it is an easy-axis antiferromagnet, it has been shown previously that strain in the crystal will cause a net moment to develop along the $c$-axis, as is observed here \cite{borovik-romanov_piezomagnetism_1960,baruchel_piezomagnetism_1988,shi_exchange_2004}. Plotting the transition temperature as a function of the (110) lattice parameter in Fig. \ref{fig:MnF2Piezo}(b), demonstrates the existence of a piezomagnetic effect in MnF$_2$ thin films. In $\sim 0.5\%$ strained MnF$_2$, the transition temperature decreases by nearly 7~K. When the strain in the MnF$_2$ thin film was fully relaxed (by growing on a (MnNi)F$_2$ graded layer), the transition temperature matched the bulk value of $66.5$ K. The graded buffer layer method used here suggests that the epitaxy-induced strain could be carefully tuned by controlling the final stoichiometry of the (MnNi)F$_2$ graded buffer layer, changing the lattice mismatch at the MnF$_2$ interface and permitting some control of the piezomagnetic behavior in thin film MnF$_2$. It is important to note here that it is difficult to differentiate magnetic moment contributions from the (MnNi)F$_2$ graded layer from the magnetic moment of the pure MnF$_2$ film itself. By its very nature, the (MnNi)F$_2$ graded layer has some thickness (less than 10~nm) that is either pure MnF$_2$ or lightly doped with NiF$_2$ that will contribute to the overall magnetization. It is unclear from these magnetic susceptibility measurements if the net moment in the relaxed MnF$_2$ film is due to interactions with the (MnNi)F$_2$ graded buffer layer or from some uncompensated strain in the MnF$_2$ thin film crystal. However, the fact that the transition temperature of the film agrees with the expected bulk value suggests that the magnetism is dominated by a relaxed, pure MnF$_2$ film. The other Mn$_x$Ni$_{1-x}$F$_2$ thin films with fixed values of $x$, discussed below, were not grown with the (MnNi)F$_2$ graded buffer layer and therefore retain some epitaxial strain.

In order to study the magnetic properties of the Mn$_x$Ni$_{1-x}$F$_2$ thin film alloys, two sets of in-plane, field-cooled (FC) thermoremanent magnetization (TRM) measurements were performed as a function of temperature. In the $c$-axis measurements, shown in Fig.~\ref{fig:MvT}(a), the samples were cooled from $T = 100$~K to $T = 4.5$~K in a small external field ($\mu_0 H=0.1$~T) applied in the film of the plane along the $c$-axis of the Mn$_x$Ni$_{1-x}$F$_2$ thin film crystal. Upon reaching $T = 4.5$~K, the external field is turned off and the magnetic moment is measured along the $c$-axis as the temperature is increased. In Fig.~\ref{fig:MvT}(b), the samples are cooled and measured in the same way, but the external field and measured moment are oriented $90^\circ$ in-plane relative to the [001] ($c$-axis) direction to measure the moment along the in-plane [1$\bar{1}$0] direction of the Mn$_x$Ni$_{1-x}$F$_2$ thin film crystal. Figure~\ref{fig:MvT}(a) shows the development of a net magnetic moment along the $c$-axis of the Mn$_x$Ni$_{1-x}$F$_2$ thin film crystal as the stoichiometry $x$ is varied. At small values of $x$, the film has little or no net moment along the [001] direction, as would be expected for a NiF$_2$ film \cite{moriya_theory_1960,shi_exchange_2004}. As the MnF$_2$ stoichiometry $x$ is increased further, a net moment develops along the [001] direction due to strain in the thin film crystal \cite{borovik-romanov_piezomagnetism_1960,baruchel_piezomagnetism_1988, shi_exchange_2004}. In contrast, Fig.~\ref{fig:MvT}(b) shows the net magnetic moment along the in-plane [1$\bar{1}$0] direction, which lies in the $a-b$ plane of the crystal. In this direction there is a large net moment even for small values of $x$ due to the DMI induced canted moment in NiF$_2$ \cite{dzyaloshinsky_thermodynamic_1958}. This net moment gradually decreases as $x$ is increased and the thin film alloy behaves more like pure MnF$_2$.

\begin{figure*}[h]
    \centering
    \includegraphics{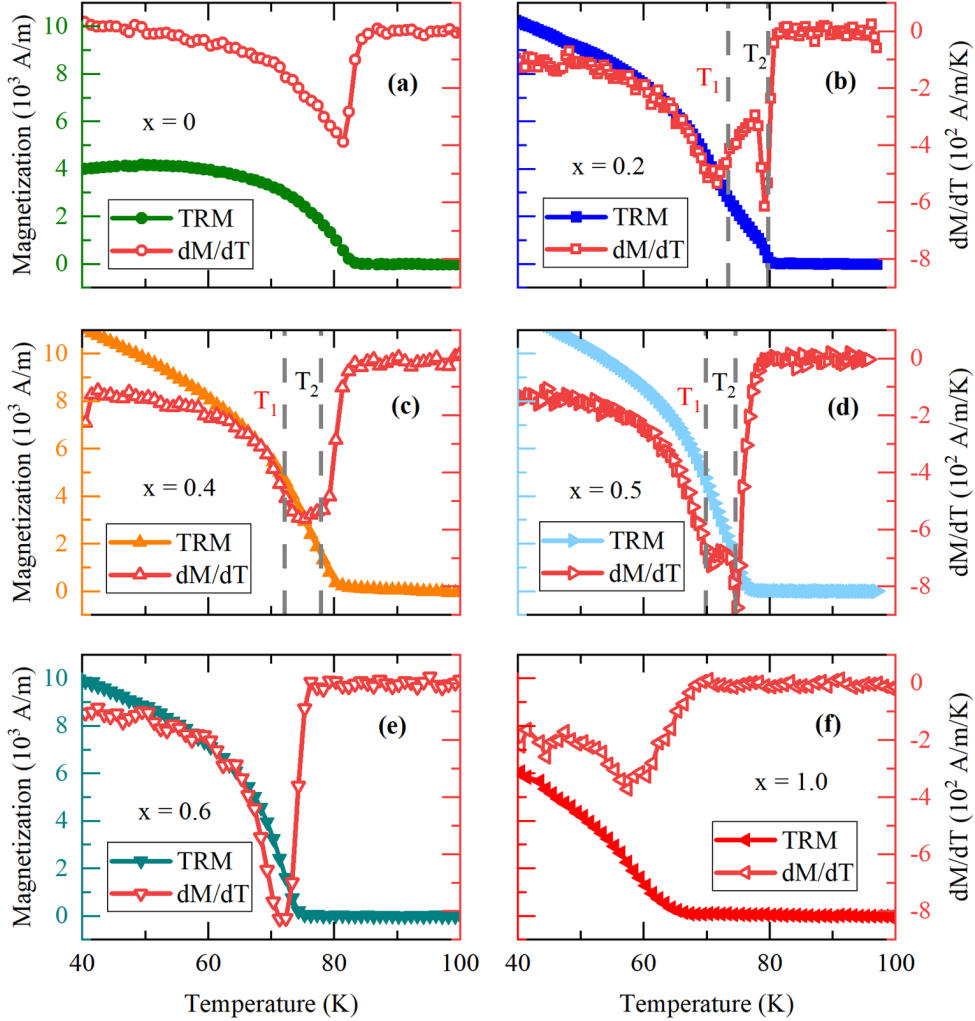}
    \caption{\label{fig:TRMdMdT} TRM and first derivative of TRM with respect to $T$ of selected samples with different values of $x$. The sample with $x = 0.2$ (b), $x = 0.4$ (c), and $x = 0.5$ (d), show two inflection points in the TRM associated with a second magnetic transition.}
    
\end{figure*}

In addition to the ordinary AF transition in Mn$_x$Ni$_{1-x}$F$_2$ thin films, magnetization measurements also show evidence of a second magnetic transition along the [1$\bar{1}$0] direction in some Mn$_x$Ni$_{1-x}$F$_2$ samples. Figure \ref{fig:TRMdMdT} shows TRM and the first derivative of the TRM as a function of temperature for several stoichiometries of Mn$_x$Ni$_{1-x}$F$_2$ films. Films shown in Fig.~\ref{fig:TRMdMdT}(a-d) are measured along the [1$\bar{1}$0] direction, while those shown in Fig.~\ref{fig:TRMdMdT}(e,f) are measured along the $c$-axis. A second magnetic transition can be identified by an inflection in the magnetization as a function of temperature, and is easily distinguished in the first derivative of the magnetization, as shown in Fig.~\ref{fig:TRMdMdT}(b), where two magnetic transitions are labeled. The ordinary AF transition is labeled as $T_2$, and the additional emergent phase is labeled as $T_1$.  

\section{Discussion}
The magnetization data can be understood in terms of a MFT approach similar to the one used to understand the Fe$_x$Ni$_{1-x}$F$_2$ system \cite{perez_phase_2015}. First consider the spin Hamiltonian  \cite{moriya_theory_1960}
\begin{equation}
    H = \sum_{i=1}^{} \sum_{j=i+1}^{} J_{ij} \mathbf{S}_i \cdot \mathbf{S}_j + D \sum_{i}^{} (S_i^z)^2  + E\left[\sum_{i}^{}(S_{ix}^2 - S_{iy}^2) - \sum_{j}^{}(S_{jx}^2 - S_{jy}^2)\right],
\end{equation}

where $J_{ij}$ is the next nearest neighbor exchange energy between spins at lattice sites $i$ and $j$, $D$ is the single-ion magnetic anisotropy energy, and $E$ is an antisymmetric exchange energy that cants moments in the $x$-$y$ plane. For the rutile structure, the $z$-direction coincides with the $c$-axis of the crystal. The known values of the spin, exchange and anisotropy energies for MnF$_2$, NiF$_2$, and FeF$_2$ are given in Table~\ref{tab:ExchangeEnergies}.

\begin{table}[h]
    \caption{\label{tab:ExchangeEnergies} Spin $S$ of the transition metal ion and magnetic exchange energies $J$, single-ion anisotropy energies $D$, and the rhombic (DM) anisotropy energy $E$ of MnF$_2$, NiF$_2$, and FeF$_2$  bulk crystals \cite{barak_magnetic_1978, hutchings_neutron_1970, hutchings_spin_1970}. The mean-field value of $J\langle S^2\rangle=JS(S+1)/3$, which is proportional to the mean-field N\'eel temperature,  is also included for reference. Energy units are in meV.}

\begin{ruledtabular}
\begin{tabular}{cccccc}

    Material & $S$ & $J$ &$JS(S+1)/3$& $D$ & $E$ \\
    \hline
    MnF$_2$ & 5/2 & 0.304 &0.887 & -0.096 & 0 \\
    NiF$_2$ & 1 & 1.719 &1.146 & 0.541 & 0.205 \\
    FeF$_2$ & 2 & 0.451 & 0.902& -0.801 & 0 \\

\end{tabular}
\end{ruledtabular}

\end{table}

\begin{figure*}[h]
    \centering
    \includegraphics{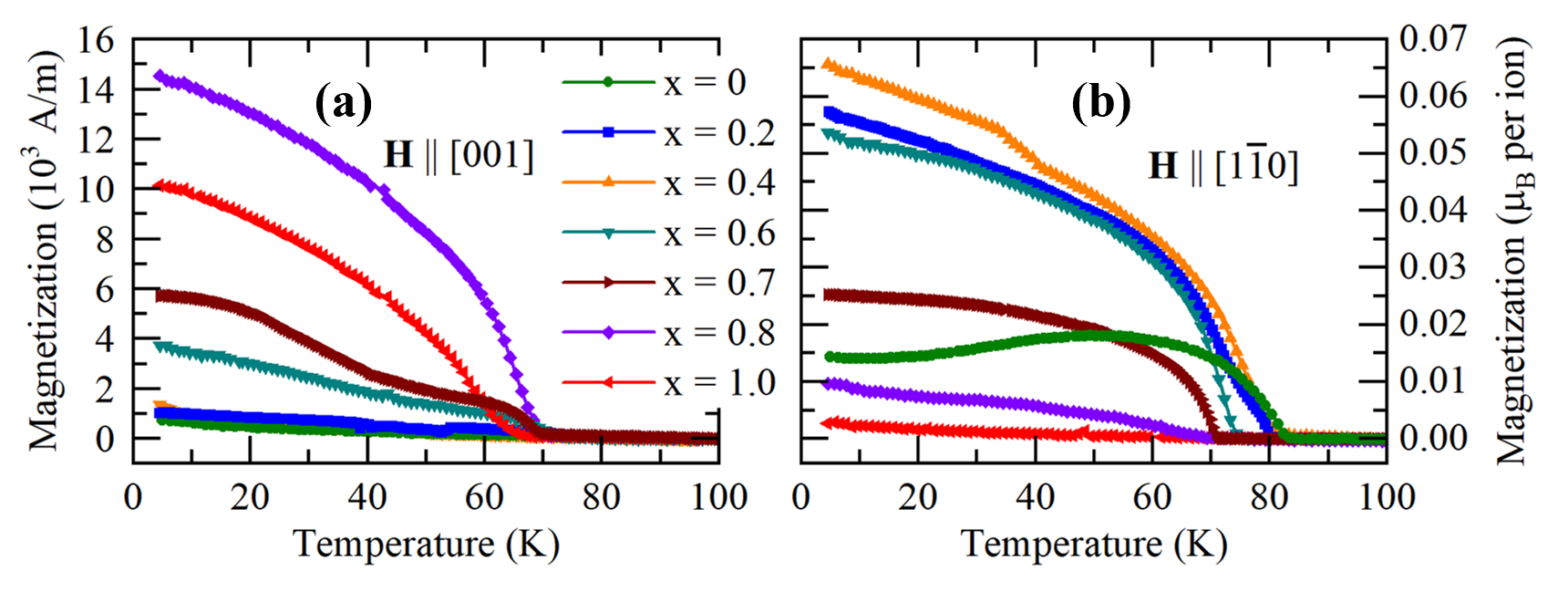}
    \caption{\label{fig:MvT} TRM as a function of temperature of Mn$_x$Ni$_{1-x}$F$_2$ thin films (a) measured along the $[001]$ ($c$-axis) and (b) along the $[1\bar{1}0]$ in-plane crystallographic directions of the samples, both of which are in the plane of the samples. The cooling external field $\mu_0 H=0.1$~T was applied along the direction of measurement.}
    
\end{figure*}

Previous studies on Fe$_{x}$Ni$_{1-x}$F$_2$ thin films have observed a similar emergent magnetic phase in magnetization measurements as a function of temperature \cite{perez_phase_2015}. Neutron diffraction measurements of the  magnetic (100) and (001) peaks showed that this state is neither the uniaxial ordering of FeF$_2$ (similar to that of MnF$_2$) nor the planar ordering of NiF$_2$ \cite{perez_phase_2015}. These observations indicate that the emergent phase could consist of a magnetic glassy state \cite{perez_phase_2015}, or, another possibility that we propose here, a helical or skyrmion phase. A skyrmion phase is hypothetically possible in this system because NiF$_2$ is known to exhibit DMI, an antisymmetric or anisotropic exchange that tends to cant magnetic moments out of antiparallel alignment and is an important ingredient in the stabilization of chiral spin textures \cite{zhang_antiferromagnetic_2016,kent_creation_2021,sergienko_role_2006}.
Verifying the latter hypothesis of a skyrmion phase requires further experimentation beyond the scope of this paper, but it could be investigated in the future with neutron scattering measurements, to check for the formation of a skyrmion lattice, or by Raman scattering measurements to check for additional magnon modes associated with either the skyrmion phase or a spin glass phase.\cite{muhlbauer_skyrmion_2009, venugopalan_raman_1981, rotaru_spin-glass_2009}.

\begin{figure}[h]
    \centering
    \includegraphics{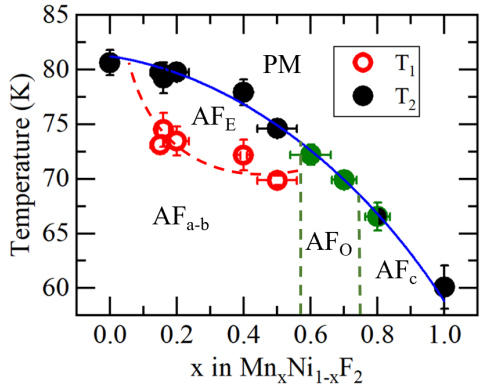}
    \caption{\label{fig:PhaseDiagram} Magnetic phase diagram of Mn$_x$Ni$_{1-x}$F$_2$ thin films divided into five regions for each phase: PM for the paramagnetic phase, AF$_\text{a-b}$ for antiferromagnetic ordering in the $a-b$ plane, AF$_\text{E}$ for the emergent magnetic phase, AF$_\text{O}$ for the oblique antiferromagnetic phase, and AF$_\text{c}$ for antiferromagnetic ordering along the $c$-axis. Samples exhibiting oblique AF order (that is, having a strong signal with $\vec{H}$ applied both parallel and perpendicular to the $c$-axis, per the data in Fig.~\ref{fig:MvT}) are colored green. The solid blue curve represents a fit to MFT. The green dashed lines indicate the region where oblique AF order is predicted to exist from MFT using the parameters given in Table~\ref{tab:ModelEnergies}. The red dashed curve is a guide to the eye approximating the emergent magnetic phase boundary from TRM measurements.}
    
\end{figure}

From the TRM measurements of the Mn$_x$Ni$_{1-x}$F$_2$ samples, we constructed the magnetic phase diagram shown in Fig.~\ref{fig:PhaseDiagram}. The solid blue curve is a fit to a MFT expression for a mixed system~\cite{wertheim_mathrmmn2-f-mathrmfe2_1969}, 
\begin{equation}\label{TNeq}
        T_N(x) = \left[p_AT_A+p_BT_B\right]/2 +\left(\frac{1}{4}[p_AT_A+p_BT)_B]^2 + p_Ap_B(T_{AB}^2 - T_AT_B)\right)^{1/2},
\end{equation} 
where $T_A$ and $T_B$ are the transition temperatures of the pure MnF$_2$ and NiF$_2$ systems, respectively, and $p_A$ and $p_B$ represent the relative stoichiometries of MnF$_2$ and NiF$_2$. Fitting the measured transition temperatures to this equation yields a value for $T_{AB}$ which can then be used to calculate the exchange integral $J_{AB}$ between the elements of the mixed system, according to the expression \cite{wertheim_mathrmmn2-f-mathrmfe2_1969}, 
\begin{equation}\label{Jeq}
    |J_{AB}| = \frac{3k_BT_{AB}}{16[S_A(S_A+1)S_B(S_B+1)]^{1/2}},
\end{equation}
where $S_A$ and $S_B$ are the spin values for the magnetic elements. This expression takes into account only the antiferromagnetic exchange between next-nearest-neighbors (between center and corner spins in the rutile structure), which is a reasonable simplification to make for MnF$_2$ and NiF$_2$ because the omitted nearest-neighbor coupling is nearly 10 times smaller than the next-nearest-neighbor coupling \cite{barak_magnetic_1978,hutchings_neutron_1970}. From Eq.~\ref{Jeq}, the exchange constant between Mn and Ni ions on opposite sublattices in our films is calculated to be $J_{\text{Mn-Ni, film}} = 0.305 \pm 0.003$~meV, compared to experimentally determined values of $J_{\text{Mn-Mn, bulk}} = 0.304 \pm 0.002$~meV in bulk MnF$_2$ \cite{nikotin_magnon_1969} and $J_{\text{Ni-Ni, bulk}} = 1.719 \pm 0.045$~meV in bulk NiF$_2$ \cite{hutchings_neutron_1970}. Assuming that the transition temperatures in the films are due to a modified value of the exchange constants resulting from strain, the values of the exchange constants in our samples are approximately $J_{\text{Mn-Mn, film}} = 0.274$~meV and  $J_{\text{Ni-Ni, film}} = 1.897$~meV. It is the exchange constant $J_{\text{Mn-Ni}}$ that is primarily responsible for the shape of the paramagnetic-AF transition in the phase diagram Fig.~\ref{fig:PhaseDiagram}. 

At low Mn stoichiometries ( $x < 0.6$ ) and below the AF transition temperature $T_2$, the magnetic moments order antiferromagnetically in the $a$-$b$ plane (AF$_\text{a-b}$), similarly to how they order in NiF$_2$. The AF$_\text{a-b}$ phase exists over such a large range of stoichiometries because of the difference in magnetic anisotropy energy $D$ between the Ni and Mn ions, with $D_{\text{NiF$_2$}}$ being more than 5 times larger than $D_{\text{MnF$_2$}}$. Within this range of AF$_\text{a-b}$ magnetic ordering, the emergent magnetic phase (AF$_\text{E}$) develops in the temperature range between $T_1$ and $T_2$, as indicated by the red dashed curve in Fig.~\ref{fig:PhaseDiagram}. As Mn stoichiometry is increased, the AF ordering enters an oblique AF phase (AF$_\text{O}$), where competition between the mutually orthogonal magnetic anisotropies of MnF$_2$ and NiF$_2$ causes the N\'eel vector to point along some angle $\theta$ between $0^\circ$ and $90^\circ$ with respect to the $a$-$b$ plane of the crystal. Samples which have strong TRM along both the $c$-axis and perpendicular to it are samples which have the oblique phase, that is, samples with $x=0.6$ and $x=0.7$ in Fig.~\ref{fig:MvT}. The sample with $x = 0.8$ is close to the AF$_\text{O}$ phase but the small TRM in the $a$-$b$ plane relative to the TRM along the $c$ axis leads us to conclude that this sample lies just outside the boundary of the AF$_\text{O}$ phase. Beyond $x = 0.8$, the system transitions into the uniaxial AF state ordering along the $c$-axis (AF$_\text{c}$) as in pure MnF$_2$. 

The oblique AF phase can be described theoretically by MFT as described in Refs.\ \cite{matsubara_mixture_1977, perez_phase_2015}. The angles $\theta_A$ and $\theta_B$ that the ions $A$ and $B$ make with respect to the $c$-axis are given by the system of equations
\begin{subequations}
\label{tan_eqns}
\begin{eqnarray}
     \tan\theta_A = \frac{z ( J_{AA} p_A S_A \sin{\theta_A} + J_{AB} p_B S_B \sin\theta_B)} {z ( J_{AA} p_A S_A \cos\theta_A + J_{AB} p_B S_B \cos\theta_B ) - 2 D_A S_A \cos\theta_A}
     \\ \nonumber \\
     \tan\theta_B = \frac{z ( J_{BB} p_B S_B \sin\theta_B + J_{AB} p_A S_A \sin\theta_A)} {z ( J_{BB} p_B S_B \cos\theta_B + J_{AB} p_A S_A \cos\theta_A ) - 2 D_B S_B \cos\theta_B}
\end{eqnarray}
\end{subequations}
where $z$ is the number of next-nearest neighbors in the lattice. For the rutile crystal structure $z=8$.

This general model of the easy axis for two anisotropic antiferromagnets successfully explains the oblique AF phase of Fe$_x$Ni$_{1-x}$F$_2$ from reference~\cite{perez_phase_2015}, predicting a stoichiometric region of $0.09 \leq x \leq 0.21$. Using our experimentally-determined values for the exchange and anisotropy constants for the Mn$_x$Ni$_{1-x}$F$_2$ system in this model predicts the existence of an oblique AF region in the stoichiometric region $0.40 \leq x \leq 0.58$, which is different from the experimentally observed oblique AF region of approximately $0.6 \leq x \leq 0.8$. One potentially important factor that is not captured by the MFT approximation is the unusual strain observed in the thin film MnF$_2$, where the crystal lattice is compressed along all three crystallographic axes. It is known that changes in the lattice spacing and unit cell volume will affect the magnetic exchange energy $J$, \cite{bloch_103_1966, shi_exchange_2004}, but the location of the oblique phase is not very sensitive to the exchange constants, per our mean field calculations. On the other hand, the stoichiometric region of oblique AF order is sensitive to changes in the anisotropy energy $D$. Decreasing the magnitude of the Mn anisotropy energy has the effect of shifting the AF$_\text{O}$-AF$_\text{c}$ phase transition to larger $x$, while increasing the magnitude of Ni anisotropy energy has the effect of shifting the AF$_\text{a-b}$-AF$_\text{O}$ phase transition to larger $x$. Increasing the overall absolute value of anisotropy energy in the system has the effect of increasing the range of the oblique phase in $x$. By decreasing 
the Mn anisotropy energy to $D_{\text{MnF}_2} = -0.06$~meV and increasing the Ni anisotropy energy to $D_{\text{NiF}_2} = 0.74$~meV, MFT predicts an oblique AF phase in the region $0.58 \leq x \leq 0.75$ at $T=0$, which agrees with the observed phase. It is possible that the compressive strain observed in thin film MnF$_2$ also affects the structure of the Mn$_x$Ni$_{1-x}$F$_2$ alloy thin films as $x$ increases and that this transition changes the anisotropy energy of the constituent Mn$^{2+}$ ions. One possible reason is that in MnF$_2$ the single-ion anisotropy is a result primarily of dipole-dipole interactions because the orbital angular momentum of  the ground state Mn$^{2+}$ is zero \cite{keffer_anisotropy_1952}, so changing the lattice parameters of the unit cell could have a significant effect on the magnitude of the single-ion anisotropy. A summary of the parameters used to reproduce our experimental data are shown in Table~\ref{tab:ModelEnergies}.
\begin{table}[h]
    \caption{\label{tab:ModelEnergies} Symbols, meaning, and values used to reproduce our experimental data from Eqs.~\ref{TNeq}, \ref{Jeq}, and \ref{tan_eqns}.}

\begin{ruledtabular}
\begin{tabular}{ccc}
Symbol&Meaning&Value\\
     \hline
    $S_A$&$S_{\text{MnF}_2}$& 5/2 \\
    $S_B$&$S_{\text{NiF}_2}$& 1  \\
    $J_{AA}$&$J_{\text{MnMn}}$&0.274 meV\\
$J_{BB}$&$J_{\text{NiNi}}$&1.897 meV\\
$J_{AB}$&$J_{\text{MnNi}}$&0.305 meV\\
$D_A$&$D_{\text{Mn}}$&-0.06 meV\\
$D_B$&$D_{\text{Ni}}$&0.74 meV\\
$p_A$&$x$&0-1 range\\
$p_B$&$1-x$&1-0 range
\end{tabular}
\end{ruledtabular}

\end{table}

The Mn$_x$Ni$_{1-x}$F$_2$ system can be compared to a similar system, Fe$_x$Ni$_{1-x}$F$_2$, to obtain some insight into the effects that the magnetic energy parameters have on the phase diagram of the system \cite{perez_phase_2015}. FeF$_2$ has the same rutile crystal structure and $c$-axis AF order as MnF$_2$, but the magnetic anisotropy energy is nearly 10 times larger than in MnF$_2$, as shown in Table~\ref{tab:ExchangeEnergies}. Note that the effective exchange interaction, proportional to $JS(S+1)/3$, is similar in MnF$_2$, NiF$_2$, and FeF$_2$, and therefore the overwhelming difference between the three systems is the single-ion anisotropy. This makes comparison between Mn$_x$Ni$_{1-x}$F$_2$, where the anisotropy is small, and Fe$_x$Ni$_{1-x}$F$_2$, where the anisotropy is large, enlightening because it illustrates the large effect that the single-ion anisotropy has in modifying the phase diagram. Specifically, the magnitude of the single-ion anisotropy energy appears to play a major role in the size of the oblique AF phase with respect to the stoichiometry. In Fe$_x$Ni$_{1-x}$F$_2$, the oblique phase is relatively small, while in Mn$_x$Ni$_{1-x}$F$_2$, with an order of magnitude smaller single-ion anisotropy, the oblique AF phase persists over a large range of stoichiometry. This behavior is both predicted by MFT and experimentally observed in magnetic susceptibility measurements of the two systems. 

It is also interesting to note that as the oblique phase grows in phase space in Mn$_x$Ni$_{1-x}$F$_2$, the emergent phase shrinks in phase space relative to Fe$_x$Ni$_{1-x}$F$_2$ \cite{perez_phase_2015}. This implies some relationship between the two magnetic states, further suggested by the fact that in both systems, Mn$_x$Ni$_{1-x}$F$_2$ and Fe$_x$Ni$_{1-x}$F$_2$, there exists a tricritical point between the emergent, oblique, and anisotropic AF phases.

\section{Conclusions}
Here we have demonstrated how an antiferromagnetic system composed of two species with competing single-ion anisotropies, embodied by Mn$_x$Ni$_{1-x}$F$_2$ alloy thin films grown via MBE, has a rich magnetic phase diagram. The Mn$_x$Ni$_{1-x}$F$_2$ thin film alloys retain their antiferromagnetic ordering as the magnetic transition temperature and lattice parameters vary with changing stoichiometry. At $ x = 1.0$, our MnF$_2$ thin films are strained in all three directions due to epitaxial growth on MgF$_2$ substrates, and as a result the AF transition temperature is reduced by 7 K. 

Using magnetization measurements of the Mn$_x$Ni$_{1-x}$F$_2$ thin film alloys along their two in plane directions, [1$\bar{1}$0] (in the $a$-$b$ plane) and [001] (along the $c$-axis), a magnetic phase diagram was constructed. MFT fits to the antiferromagnetic transition temperature allow us to calculate the magnetic exchange energy between the Mn and Ni ions to be $J_{\text{Mn-Ni}} = 0.305 \pm 0.003 \, \text{meV}$. A MFT approximation was used to predict the existence of an oblique AF phase observed in the experimental magnetization measurements. Agreement with the experimental data also requires a decrease in the single-ion anisotropy in MnF$_2$ with respect to the bulk material, possibly as a result of the strain induced by the epitaxial growth of the films on the MgF$_2$. The oblique AF phase shares a tricritical point with the NiF$_2$-like anisotropic AF phase and an emergent magnetic phase with unidentified structure.  This emergent magnetic phase could be a magnetic glassy phase or a helical (or possibly skyrmion) phase. Further experimental and theoretical work needs to be performed to identify the structure of the emergent phase in this system.

\begin{acknowledgments}
This work was supported in part by the Air Force MURI program, grant \# FA9550-19-1-0307. Work on sample growth by N.\ Hald was performed at UC Santa Cruz and was supported by the UCSC REU program through the National Science Foundation, grant \# 1950907.
\end{acknowledgments}


\clearpage

\bibliography{NiMnF2_alloy_paper}

\begin{thebibliography}{32}%
\makeatletter
\providecommand \@ifxundefined [1]{%
 \@ifx{#1\undefined}
}%
\providecommand \@ifnum [1]{%
 \ifnum #1\expandafter \@firstoftwo
 \else \expandafter \@secondoftwo
 \fi
}%
\providecommand \@ifx [1]{%
 \ifx #1\expandafter \@firstoftwo
 \else \expandafter \@secondoftwo
 \fi
}%
\providecommand \natexlab [1]{#1}%
\providecommand \enquote  [1]{``#1''}%
\providecommand \bibnamefont  [1]{#1}%
\providecommand \bibfnamefont [1]{#1}%
\providecommand \citenamefont [1]{#1}%
\providecommand \href@noop [0]{\@secondoftwo}%
\providecommand \href [0]{\begingroup \@sanitize@url \@href}%
\providecommand \@href[1]{\@@startlink{#1}\@@href}%
\providecommand \@@href[1]{\endgroup#1\@@endlink}%
\providecommand \@sanitize@url [0]{\catcode `\\12\catcode `\$12\catcode
  `\&12\catcode `\#12\catcode `\^12\catcode `\_12\catcode `\%12\relax}%
\providecommand \@@startlink[1]{}%
\providecommand \@@endlink[0]{}%
\providecommand \url  [0]{\begingroup\@sanitize@url \@url }%
\providecommand \@url [1]{\endgroup\@href {#1}{\urlprefix }}%
\providecommand \urlprefix  [0]{URL }%
\providecommand \Eprint [0]{\href }%
\providecommand \doibase [0]{https://doi.org/}%
\providecommand \selectlanguage [0]{\@gobble}%
\providecommand \bibinfo  [0]{\@secondoftwo}%
\providecommand \bibfield  [0]{\@secondoftwo}%
\providecommand \translation [1]{[#1]}%
\providecommand \BibitemOpen [0]{}%
\providecommand \bibitemStop [0]{}%
\providecommand \bibitemNoStop [0]{.\EOS\space}%
\providecommand \EOS [0]{\spacefactor3000\relax}%
\providecommand \BibitemShut  [1]{\csname bibitem#1\endcsname}%
\let\auto@bib@innerbib\@empty
\bibitem [{\citenamefont {Stout}\ and\ \citenamefont
  {Reed}(1954)}]{stout_crystal_1954}%
  \BibitemOpen
  \bibfield  {author} {\bibinfo {author} {\bibfnamefont {J.~W.}\ \bibnamefont
  {Stout}}\ and\ \bibinfo {author} {\bibfnamefont {S.~A.}\ \bibnamefont
  {Reed}},\ }\href {https://doi.org/10.1021/ja01650a005} {\bibfield  {journal}
  {\bibinfo  {journal} {Journal of the American Chemical Society}\ }\textbf
  {\bibinfo {volume} {76}},\ \bibinfo {pages} {5279} (\bibinfo {year}
  {1954})},\ \bibinfo {note} {publisher: American Chemical Society}\BibitemShut
  {NoStop}%
\bibitem [{\citenamefont
  {Dzyaloshinsky}(1958)}]{dzyaloshinsky_thermodynamic_1958}%
  \BibitemOpen
  \bibfield  {author} {\bibinfo {author} {\bibfnamefont {I.}~\bibnamefont
  {Dzyaloshinsky}},\ }\href {https://doi.org/10.1016/0022-3697(58)90076-3}
  {\bibfield  {journal} {\bibinfo  {journal} {Journal of Physics and Chemistry
  of Solids}\ }\textbf {\bibinfo {volume} {4}},\ \bibinfo {pages} {241}
  (\bibinfo {year} {1958})}\BibitemShut {NoStop}%
\bibitem [{\citenamefont {Moriya}(1960)}]{moriya_theory_1960}%
  \BibitemOpen
  \bibfield  {author} {\bibinfo {author} {\bibfnamefont {T.}~\bibnamefont
  {Moriya}},\ }\href {https://doi.org/10.1103/PhysRev.117.635} {\bibfield
  {journal} {\bibinfo  {journal} {Physical Review}\ }\textbf {\bibinfo {volume}
  {117}},\ \bibinfo {pages} {635} (\bibinfo {year} {1960})},\ \bibinfo {note}
  {publisher: American Physical Society}\BibitemShut {NoStop}%
\bibitem [{\citenamefont {Borovik-Romanov}\ \emph {et~al.}(1973)\citenamefont
  {Borovik-Romanov}, \citenamefont {Bazhan},\ and\ \citenamefont
  {Kreines}}]{borovik-romanov_weak_1973}%
  \BibitemOpen
  \bibfield  {author} {\bibinfo {author} {\bibfnamefont {A.}~\bibnamefont
  {Borovik-Romanov}}, \bibinfo {author} {\bibfnamefont {A.}~\bibnamefont
  {Bazhan}},\ and\ \bibinfo {author} {\bibfnamefont {N.}~\bibnamefont
  {Kreines}},\ }\href {http://jetp.ras.ru/cgi-bin/dn/e_037_04_0695.pdf}
  {\bibfield  {journal} {\bibinfo  {journal} {Journal of Experimental and
  Theoretical Physics}\ }\textbf {\bibinfo {volume} {37}},\ \bibinfo {pages}
  {695} (\bibinfo {year} {1973})}\BibitemShut {NoStop}%
\bibitem [{\citenamefont {Emori}\ \emph {et~al.}(2013)\citenamefont {Emori},
  \citenamefont {Bauer}, \citenamefont {Ahn}, \citenamefont {Martinez},\ and\
  \citenamefont {Beach}}]{emori_current-driven_2013}%
  \BibitemOpen
  \bibfield  {author} {\bibinfo {author} {\bibfnamefont {S.}~\bibnamefont
  {Emori}}, \bibinfo {author} {\bibfnamefont {U.}~\bibnamefont {Bauer}},
  \bibinfo {author} {\bibfnamefont {S.-M.}\ \bibnamefont {Ahn}}, \bibinfo
  {author} {\bibfnamefont {E.}~\bibnamefont {Martinez}},\ and\ \bibinfo
  {author} {\bibfnamefont {G.~S.~D.}\ \bibnamefont {Beach}},\ }\href
  {https://doi.org/10.1038/nmat3675} {\bibfield  {journal} {\bibinfo  {journal}
  {Nature Materials}\ }\textbf {\bibinfo {volume} {12}},\ \bibinfo {pages}
  {611} (\bibinfo {year} {2013})},\ \bibinfo {note} {number: 7 Publisher:
  Nature Publishing Group}\BibitemShut {NoStop}%
\bibitem [{\citenamefont {Barker}\ and\ \citenamefont
  {Tretiakov}(2016)}]{barker_static_2016}%
  \BibitemOpen
  \bibfield  {author} {\bibinfo {author} {\bibfnamefont {J.}~\bibnamefont
  {Barker}}\ and\ \bibinfo {author} {\bibfnamefont {O.~A.}\ \bibnamefont
  {Tretiakov}},\ }\href {https://doi.org/10.1103/PhysRevLett.116.147203}
  {\bibfield  {journal} {\bibinfo  {journal} {Physical Review Letters}\
  }\textbf {\bibinfo {volume} {116}},\ \bibinfo {pages} {147203} (\bibinfo
  {year} {2016})},\ \bibinfo {note} {publisher: American Physical
  Society}\BibitemShut {NoStop}%
\bibitem [{\citenamefont {Zhang}\ \emph {et~al.}(2016)\citenamefont {Zhang},
  \citenamefont {Zhou},\ and\ \citenamefont
  {Ezawa}}]{zhang_antiferromagnetic_2016}%
  \BibitemOpen
  \bibfield  {author} {\bibinfo {author} {\bibfnamefont {X.}~\bibnamefont
  {Zhang}}, \bibinfo {author} {\bibfnamefont {Y.}~\bibnamefont {Zhou}},\ and\
  \bibinfo {author} {\bibfnamefont {M.}~\bibnamefont {Ezawa}},\ }\href
  {https://doi.org/10.1038/srep24795} {\bibfield  {journal} {\bibinfo
  {journal} {Scientific Reports}\ }\textbf {\bibinfo {volume} {6}},\ \bibinfo
  {pages} {24795} (\bibinfo {year} {2016})},\ \bibinfo {note} {number: 1
  Publisher: Nature Publishing Group}\BibitemShut {NoStop}%
\bibitem [{\citenamefont {Kent}\ \emph {et~al.}(2021)\citenamefont {Kent},
  \citenamefont {Reynolds}, \citenamefont {Raftrey}, \citenamefont {Campbell},
  \citenamefont {Virasawmy}, \citenamefont {Dhuey}, \citenamefont {Chopdekar},
  \citenamefont {Hierro-Rodriguez}, \citenamefont {Sorrentino}, \citenamefont
  {Pereiro}, \citenamefont {Ferrer}, \citenamefont {Hellman}, \citenamefont
  {Sutcliffe},\ and\ \citenamefont {Fischer}}]{kent_creation_2021}%
  \BibitemOpen
  \bibfield  {author} {\bibinfo {author} {\bibfnamefont {N.}~\bibnamefont
  {Kent}}, \bibinfo {author} {\bibfnamefont {N.}~\bibnamefont {Reynolds}},
  \bibinfo {author} {\bibfnamefont {D.}~\bibnamefont {Raftrey}}, \bibinfo
  {author} {\bibfnamefont {I.~T.~G.}\ \bibnamefont {Campbell}}, \bibinfo
  {author} {\bibfnamefont {S.}~\bibnamefont {Virasawmy}}, \bibinfo {author}
  {\bibfnamefont {S.}~\bibnamefont {Dhuey}}, \bibinfo {author} {\bibfnamefont
  {R.~V.}\ \bibnamefont {Chopdekar}}, \bibinfo {author} {\bibfnamefont
  {A.}~\bibnamefont {Hierro-Rodriguez}}, \bibinfo {author} {\bibfnamefont
  {A.}~\bibnamefont {Sorrentino}}, \bibinfo {author} {\bibfnamefont
  {E.}~\bibnamefont {Pereiro}}, \bibinfo {author} {\bibfnamefont
  {S.}~\bibnamefont {Ferrer}}, \bibinfo {author} {\bibfnamefont
  {F.}~\bibnamefont {Hellman}}, \bibinfo {author} {\bibfnamefont
  {P.}~\bibnamefont {Sutcliffe}},\ and\ \bibinfo {author} {\bibfnamefont
  {P.}~\bibnamefont {Fischer}},\ }\href
  {https://doi.org/10.1038/s41467-021-21846-5} {\bibfield  {journal} {\bibinfo
  {journal} {Nature Communications}\ }\textbf {\bibinfo {volume} {12}},\
  \bibinfo {pages} {1562} (\bibinfo {year} {2021})},\ \bibinfo {note} {number:
  1 Publisher: Nature Publishing Group}\BibitemShut {NoStop}%
\bibitem [{\citenamefont {Rybakov}\ \emph {et~al.}(2022)\citenamefont
  {Rybakov}, \citenamefont {Kiselev}, \citenamefont {Borisov}, \citenamefont
  {Döring}, \citenamefont {Melcher},\ and\ \citenamefont
  {Blügel}}]{rybakov_magnetic_2022}%
  \BibitemOpen
  \bibfield  {author} {\bibinfo {author} {\bibfnamefont {F.~N.}\ \bibnamefont
  {Rybakov}}, \bibinfo {author} {\bibfnamefont {N.~S.}\ \bibnamefont
  {Kiselev}}, \bibinfo {author} {\bibfnamefont {A.~B.}\ \bibnamefont
  {Borisov}}, \bibinfo {author} {\bibfnamefont {L.}~\bibnamefont {Döring}},
  \bibinfo {author} {\bibfnamefont {C.}~\bibnamefont {Melcher}},\ and\ \bibinfo
  {author} {\bibfnamefont {S.}~\bibnamefont {Blügel}},\ }\href
  {https://doi.org/10.1063/5.0099942} {\bibfield  {journal} {\bibinfo
  {journal} {APL Materials}\ }\textbf {\bibinfo {volume} {10}},\ \bibinfo
  {pages} {111113} (\bibinfo {year} {2022})},\ \bibinfo {note} {publisher:
  American Institute of Physics}\BibitemShut {NoStop}%
\bibitem [{\citenamefont {Barak}\ \emph {et~al.}(1978)\citenamefont {Barak},
  \citenamefont {Jaccarino},\ and\ \citenamefont
  {Rezende}}]{barak_magnetic_1978}%
  \BibitemOpen
  \bibfield  {author} {\bibinfo {author} {\bibfnamefont {J.}~\bibnamefont
  {Barak}}, \bibinfo {author} {\bibfnamefont {V.}~\bibnamefont {Jaccarino}},\
  and\ \bibinfo {author} {\bibfnamefont {S.~M.}\ \bibnamefont {Rezende}},\
  }\href {https://doi.org/10.1016/0304-8853(78)90087-2} {\bibfield  {journal}
  {\bibinfo  {journal} {Journal of Magnetism and Magnetic Materials}\ }\textbf
  {\bibinfo {volume} {9}},\ \bibinfo {pages} {323} (\bibinfo {year}
  {1978})}\BibitemShut {NoStop}%
\bibitem [{\citenamefont {Wu}\ \emph {et~al.}(2016)\citenamefont {Wu},
  \citenamefont {Zhang}, \citenamefont {KC}, \citenamefont {Borisov},
  \citenamefont {Pearson}, \citenamefont {Jiang}, \citenamefont {Lederman},
  \citenamefont {Hoffmann},\ and\ \citenamefont
  {Bhattacharya}}]{wu_antiferromagnetic_2016}%
  \BibitemOpen
  \bibfield  {author} {\bibinfo {author} {\bibfnamefont {S.~M.}\ \bibnamefont
  {Wu}}, \bibinfo {author} {\bibfnamefont {W.}~\bibnamefont {Zhang}}, \bibinfo
  {author} {\bibfnamefont {A.}~\bibnamefont {KC}}, \bibinfo {author}
  {\bibfnamefont {P.}~\bibnamefont {Borisov}}, \bibinfo {author} {\bibfnamefont
  {J.~E.}\ \bibnamefont {Pearson}}, \bibinfo {author} {\bibfnamefont {J.~S.}\
  \bibnamefont {Jiang}}, \bibinfo {author} {\bibfnamefont {D.}~\bibnamefont
  {Lederman}}, \bibinfo {author} {\bibfnamefont {A.}~\bibnamefont {Hoffmann}},\
  and\ \bibinfo {author} {\bibfnamefont {A.}~\bibnamefont {Bhattacharya}},\
  }\href {https://doi.org/10.1103/PhysRevLett.116.097204} {\bibfield  {journal}
  {\bibinfo  {journal} {Physical Review Letters}\ }\textbf {\bibinfo {volume}
  {116}},\ \bibinfo {pages} {097204} (\bibinfo {year} {2016})},\ \bibinfo
  {note} {publisher: American Physical Society}\BibitemShut {NoStop}%
\bibitem [{\citenamefont {Zhao}\ \emph {et~al.}(2006)\citenamefont {Zhao},
  \citenamefont {Bragas}, \citenamefont {Merlin},\ and\ \citenamefont
  {Lockwood}}]{zhao_magnon_2006}%
  \BibitemOpen
  \bibfield  {author} {\bibinfo {author} {\bibfnamefont {J.}~\bibnamefont
  {Zhao}}, \bibinfo {author} {\bibfnamefont {A.~V.}\ \bibnamefont {Bragas}},
  \bibinfo {author} {\bibfnamefont {R.}~\bibnamefont {Merlin}},\ and\ \bibinfo
  {author} {\bibfnamefont {D.~J.}\ \bibnamefont {Lockwood}},\ }\href
  {https://doi.org/10.1103/PhysRevB.73.184434} {\bibfield  {journal} {\bibinfo
  {journal} {Physical Review B}\ }\textbf {\bibinfo {volume} {73}},\ \bibinfo
  {pages} {184434} (\bibinfo {year} {2006})},\ \bibinfo {note} {publisher:
  American Physical Society}\BibitemShut {NoStop}%
\bibitem [{\citenamefont {Jacobs}(1961)}]{jacobs_spin-flopping_1961}%
  \BibitemOpen
  \bibfield  {author} {\bibinfo {author} {\bibfnamefont {I.~S.}\ \bibnamefont
  {Jacobs}},\ }\href {https://doi.org/10.1063/1.2000500} {\bibfield  {journal}
  {\bibinfo  {journal} {Journal of Applied Physics}\ }\textbf {\bibinfo
  {volume} {32}},\ \bibinfo {pages} {S61} (\bibinfo {year} {1961})}\BibitemShut
  {NoStop}%
\bibitem [{\citenamefont {Felcher}\ and\ \citenamefont
  {Kleb}(1996)}]{felcher_antiferromagnetic_1996}%
  \BibitemOpen
  \bibfield  {author} {\bibinfo {author} {\bibfnamefont {G.~P.}\ \bibnamefont
  {Felcher}}\ and\ \bibinfo {author} {\bibfnamefont {R.}~\bibnamefont {Kleb}},\
  }\href {https://doi.org/10.1209/epl/i1996-00251-7} {\bibfield  {journal}
  {\bibinfo  {journal} {Europhysics Letters (EPL)}\ }\textbf {\bibinfo {volume}
  {36}},\ \bibinfo {pages} {455} (\bibinfo {year} {1996})}\BibitemShut
  {NoStop}%
\bibitem [{\citenamefont {Perez}\ \emph {et~al.}(2015)\citenamefont {Perez},
  \citenamefont {Borisov}, \citenamefont {Johnson}, \citenamefont {Stanescu},
  \citenamefont {Trappen}, \citenamefont {Holcomb}, \citenamefont {Lederman},
  \citenamefont {Fitzsimmons}, \citenamefont {Aczel},\ and\ \citenamefont
  {Hong}}]{perez_phase_2015}%
  \BibitemOpen
  \bibfield  {author} {\bibinfo {author} {\bibfnamefont {F.~A.}\ \bibnamefont
  {Perez}}, \bibinfo {author} {\bibfnamefont {P.}~\bibnamefont {Borisov}},
  \bibinfo {author} {\bibfnamefont {T.~A.}\ \bibnamefont {Johnson}}, \bibinfo
  {author} {\bibfnamefont {T.~D.}\ \bibnamefont {Stanescu}}, \bibinfo {author}
  {\bibfnamefont {R.}~\bibnamefont {Trappen}}, \bibinfo {author} {\bibfnamefont
  {M.~B.}\ \bibnamefont {Holcomb}}, \bibinfo {author} {\bibfnamefont
  {D.}~\bibnamefont {Lederman}}, \bibinfo {author} {\bibfnamefont
  {M.}~\bibnamefont {Fitzsimmons}}, \bibinfo {author} {\bibfnamefont {A.~A.}\
  \bibnamefont {Aczel}},\ and\ \bibinfo {author} {\bibfnamefont
  {T.}~\bibnamefont {Hong}},\ }\href
  {https://doi.org/10.1103/PhysRevLett.114.097201} {\bibfield  {journal}
  {\bibinfo  {journal} {Physical Review Letters}\ }\textbf {\bibinfo {volume}
  {114}},\ \bibinfo {pages} {097201} (\bibinfo {year} {2015})},\ \bibinfo
  {note} {publisher: American Physical Society}\BibitemShut {NoStop}%
\bibitem [{\citenamefont {Hutchings}\ \emph
  {et~al.}(1970{\natexlab{a}})\citenamefont {Hutchings}, \citenamefont
  {Rainford},\ and\ \citenamefont {Guggenheim}}]{hutchings_spin_1970}%
  \BibitemOpen
  \bibfield  {author} {\bibinfo {author} {\bibfnamefont {M.~T.}\ \bibnamefont
  {Hutchings}}, \bibinfo {author} {\bibfnamefont {B.~D.}\ \bibnamefont
  {Rainford}},\ and\ \bibinfo {author} {\bibfnamefont {H.~J.}\ \bibnamefont
  {Guggenheim}},\ }\href {https://doi.org/10.1088/0022-3719/3/2/013} {\bibfield
   {journal} {\bibinfo  {journal} {Journal of Physics C: Solid State Physics}\
  }\textbf {\bibinfo {volume} {3}},\ \bibinfo {pages} {307} (\bibinfo {year}
  {1970}{\natexlab{a}})}\BibitemShut {NoStop}%
\bibitem [{\citenamefont
  {Borovik-Romanov}(1960)}]{borovik-romanov_piezomagnetism_1960}%
  \BibitemOpen
  \bibfield  {author} {\bibinfo {author} {\bibfnamefont {A.}~\bibnamefont
  {Borovik-Romanov}},\ }\href
  {http://www.jetp.ras.ru/cgi-bin/dn/e_011_04_0786.pdf} {\bibfield  {journal}
  {\bibinfo  {journal} {Soviet Physics JETP}\ }\textbf {\bibinfo {volume}
  {11}},\ \bibinfo {pages} {786} (\bibinfo {year} {1960})}\BibitemShut
  {NoStop}%
\bibitem [{\citenamefont {Baruchel}\ \emph {et~al.}(1980)\citenamefont
  {Baruchel}, \citenamefont {Schlenker},\ and\ \citenamefont
  {Barbara}}]{baruchel_180_1980}%
  \BibitemOpen
  \bibfield  {author} {\bibinfo {author} {\bibfnamefont {J.}~\bibnamefont
  {Baruchel}}, \bibinfo {author} {\bibfnamefont {M.}~\bibnamefont
  {Schlenker}},\ and\ \bibinfo {author} {\bibfnamefont {B.}~\bibnamefont
  {Barbara}},\ }\href {https://doi.org/10.1016/0304-8853(80)90388-1} {\bibfield
   {journal} {\bibinfo  {journal} {Journal of Magnetism and Magnetic
  Materials}\ }\textbf {\bibinfo {volume} {15-18}},\ \bibinfo {pages} {1510}
  (\bibinfo {year} {1980})}\BibitemShut {NoStop}%
\bibitem [{\citenamefont {Baruchel}\ \emph {et~al.}(1988)\citenamefont
  {Baruchel}, \citenamefont {Draperi}, \citenamefont {El~Kadiri}, \citenamefont
  {Fillion}, \citenamefont {Maeder}, \citenamefont {Molho},\ and\ \citenamefont
  {Porteseil}}]{baruchel_piezomagnetism_1988}%
  \BibitemOpen
  \bibfield  {author} {\bibinfo {author} {\bibfnamefont {J.}~\bibnamefont
  {Baruchel}}, \bibinfo {author} {\bibfnamefont {A.}~\bibnamefont {Draperi}},
  \bibinfo {author} {\bibfnamefont {M.}~\bibnamefont {El~Kadiri}}, \bibinfo
  {author} {\bibfnamefont {G.}~\bibnamefont {Fillion}}, \bibinfo {author}
  {\bibfnamefont {M.}~\bibnamefont {Maeder}}, \bibinfo {author} {\bibfnamefont
  {P.}~\bibnamefont {Molho}},\ and\ \bibinfo {author} {\bibfnamefont {J.~L.}\
  \bibnamefont {Porteseil}},\ }\href
  {https://doi.org/10.1051/jphyscol:19888859} {\bibfield  {journal} {\bibinfo
  {journal} {Le Journal de Physique Colloques}\ }\textbf {\bibinfo {volume}
  {49}},\ \bibinfo {pages} {C8} (\bibinfo {year} {1988})}\BibitemShut {NoStop}%
\bibitem [{\citenamefont {Baur}(1976)}]{baur_rutile-type_1976}%
  \BibitemOpen
  \bibfield  {author} {\bibinfo {author} {\bibfnamefont {W.}~\bibnamefont
  {Baur}},\ }\href {https://doi.org/10.1107/S0567740876007371} {\bibfield
  {journal} {\bibinfo  {journal} {Acta Crystallographica Section B Structural
  Crystallography and Crystal Chemistry}\ }\textbf {\bibinfo {volume} {32}},\
  \bibinfo {pages} {2200} (\bibinfo {year} {1976})}\BibitemShut {NoStop}%
\bibitem [{Van()}]{VanHaren2023}%
  \BibitemOpen
  \href@noop {} {\bibinfo {title} {{See Supplemental Material at [URL will be
  inserted by publisher] for additional discussion of XRD measurements, bulk
  lattice parameters, RHEED patterns, and derivation of oblique phase equations
  from MFT.}}}\BibitemShut {Stop}%
\bibitem [{\citenamefont {Shi}\ \emph {et~al.}(2004)\citenamefont {Shi},
  \citenamefont {Lederman}, \citenamefont {O’Donovan},\ and\ \citenamefont
  {Borchers}}]{shi_exchange_2004}%
  \BibitemOpen
  \bibfield  {author} {\bibinfo {author} {\bibfnamefont {H.}~\bibnamefont
  {Shi}}, \bibinfo {author} {\bibfnamefont {D.}~\bibnamefont {Lederman}},
  \bibinfo {author} {\bibfnamefont {K.~V.}\ \bibnamefont {O’Donovan}},\ and\
  \bibinfo {author} {\bibfnamefont {J.~A.}\ \bibnamefont {Borchers}},\ }\href
  {https://doi.org/10.1103/PhysRevB.69.214416} {\bibfield  {journal} {\bibinfo
  {journal} {Physical Review B}\ }\textbf {\bibinfo {volume} {69}},\ \bibinfo
  {pages} {214416} (\bibinfo {year} {2004})},\ \bibinfo {note} {publisher:
  American Physical Society}\BibitemShut {NoStop}%
\bibitem [{\citenamefont {Hutchings}\ \emph
  {et~al.}(1970{\natexlab{b}})\citenamefont {Hutchings}, \citenamefont
  {Thorpe}, \citenamefont {Birgeneau}, \citenamefont {Fleury},\ and\
  \citenamefont {Guggenheim}}]{hutchings_neutron_1970}%
  \BibitemOpen
  \bibfield  {author} {\bibinfo {author} {\bibfnamefont {M.~T.}\ \bibnamefont
  {Hutchings}}, \bibinfo {author} {\bibfnamefont {M.~F.}\ \bibnamefont
  {Thorpe}}, \bibinfo {author} {\bibfnamefont {R.~J.}\ \bibnamefont
  {Birgeneau}}, \bibinfo {author} {\bibfnamefont {P.~A.}\ \bibnamefont
  {Fleury}},\ and\ \bibinfo {author} {\bibfnamefont {H.~J.}\ \bibnamefont
  {Guggenheim}},\ }\href {https://doi.org/10.1103/PhysRevB.2.1362} {\bibfield
  {journal} {\bibinfo  {journal} {Physical Review B}\ }\textbf {\bibinfo
  {volume} {2}},\ \bibinfo {pages} {1362} (\bibinfo {year}
  {1970}{\natexlab{b}})},\ \bibinfo {note} {publisher: American Physical
  Society}\BibitemShut {NoStop}%
\bibitem [{\citenamefont {Sergienko}\ and\ \citenamefont
  {Dagotto}(2006)}]{sergienko_role_2006}%
  \BibitemOpen
  \bibfield  {author} {\bibinfo {author} {\bibfnamefont {I.~A.}\ \bibnamefont
  {Sergienko}}\ and\ \bibinfo {author} {\bibfnamefont {E.}~\bibnamefont
  {Dagotto}},\ }\href {https://doi.org/10.1103/PhysRevB.73.094434} {\bibfield
  {journal} {\bibinfo  {journal} {Physical Review B}\ }\textbf {\bibinfo
  {volume} {73}},\ \bibinfo {pages} {094434} (\bibinfo {year} {2006})},\
  \bibinfo {note} {publisher: American Physical Society}\BibitemShut {NoStop}%
\bibitem [{\citenamefont {Mühlbauer}\ \emph {et~al.}(2009)\citenamefont
  {Mühlbauer}, \citenamefont {Binz}, \citenamefont {Jonietz}, \citenamefont
  {Pfleiderer}, \citenamefont {Rosch}, \citenamefont {Neubauer}, \citenamefont
  {Georgii},\ and\ \citenamefont {Böni}}]{muhlbauer_skyrmion_2009}%
  \BibitemOpen
  \bibfield  {author} {\bibinfo {author} {\bibfnamefont {S.}~\bibnamefont
  {Mühlbauer}}, \bibinfo {author} {\bibfnamefont {B.}~\bibnamefont {Binz}},
  \bibinfo {author} {\bibfnamefont {F.}~\bibnamefont {Jonietz}}, \bibinfo
  {author} {\bibfnamefont {C.}~\bibnamefont {Pfleiderer}}, \bibinfo {author}
  {\bibfnamefont {A.}~\bibnamefont {Rosch}}, \bibinfo {author} {\bibfnamefont
  {A.}~\bibnamefont {Neubauer}}, \bibinfo {author} {\bibfnamefont
  {R.}~\bibnamefont {Georgii}},\ and\ \bibinfo {author} {\bibfnamefont
  {P.}~\bibnamefont {Böni}},\ }\href {https://doi.org/10.1126/science.1166767}
  {\bibfield  {journal} {\bibinfo  {journal} {Science}\ }\textbf {\bibinfo
  {volume} {323}},\ \bibinfo {pages} {915} (\bibinfo {year} {2009})},\ \bibinfo
  {note} {publisher: American Association for the Advancement of
  Science}\BibitemShut {NoStop}%
\bibitem [{\citenamefont {Venugopalan}\ \emph {et~al.}(1981)\citenamefont
  {Venugopalan}, \citenamefont {Petrou}, \citenamefont {Galazka},\ and\
  \citenamefont {Ramdas}}]{venugopalan_raman_1981}%
  \BibitemOpen
  \bibfield  {author} {\bibinfo {author} {\bibfnamefont {S.}~\bibnamefont
  {Venugopalan}}, \bibinfo {author} {\bibfnamefont {A.}~\bibnamefont {Petrou}},
  \bibinfo {author} {\bibfnamefont {R.~R.}\ \bibnamefont {Galazka}},\ and\
  \bibinfo {author} {\bibfnamefont {A.~K.}\ \bibnamefont {Ramdas}},\ }\href
  {https://doi.org/10.1016/0038-1098(81)90259-3} {\bibfield  {journal}
  {\bibinfo  {journal} {Solid State Communications}\ }\textbf {\bibinfo
  {volume} {38}},\ \bibinfo {pages} {365} (\bibinfo {year} {1981})}\BibitemShut
  {NoStop}%
\bibitem [{\citenamefont {Rotaru}\ \emph {et~al.}(2009)\citenamefont {Rotaru},
  \citenamefont {Roessli}, \citenamefont {Amato}, \citenamefont {Gvasaliya},
  \citenamefont {Mudry}, \citenamefont {Lushnikov},\ and\ \citenamefont
  {Shaplygina}}]{rotaru_spin-glass_2009}%
  \BibitemOpen
  \bibfield  {author} {\bibinfo {author} {\bibfnamefont {G.~M.}\ \bibnamefont
  {Rotaru}}, \bibinfo {author} {\bibfnamefont {B.}~\bibnamefont {Roessli}},
  \bibinfo {author} {\bibfnamefont {A.}~\bibnamefont {Amato}}, \bibinfo
  {author} {\bibfnamefont {S.~N.}\ \bibnamefont {Gvasaliya}}, \bibinfo {author}
  {\bibfnamefont {C.}~\bibnamefont {Mudry}}, \bibinfo {author} {\bibfnamefont
  {S.~G.}\ \bibnamefont {Lushnikov}},\ and\ \bibinfo {author} {\bibfnamefont
  {T.~A.}\ \bibnamefont {Shaplygina}},\ }\href
  {https://doi.org/10.1103/PhysRevB.79.184430} {\bibfield  {journal} {\bibinfo
  {journal} {Physical Review B}\ }\textbf {\bibinfo {volume} {79}},\ \bibinfo
  {pages} {184430} (\bibinfo {year} {2009})},\ \bibinfo {note} {publisher:
  American Physical Society}\BibitemShut {NoStop}%
\bibitem [{\citenamefont {Wertheim}\ \emph {et~al.}(1969)\citenamefont
  {Wertheim}, \citenamefont {Guggenheim}, \citenamefont {Butler},\ and\
  \citenamefont {Jaccarino}}]{wertheim_mathrmmn2-f-mathrmfe2_1969}%
  \BibitemOpen
  \bibfield  {author} {\bibinfo {author} {\bibfnamefont {G.~K.}\ \bibnamefont
  {Wertheim}}, \bibinfo {author} {\bibfnamefont {H.~J.}\ \bibnamefont
  {Guggenheim}}, \bibinfo {author} {\bibfnamefont {M.}~\bibnamefont {Butler}},\
  and\ \bibinfo {author} {\bibfnamefont {V.}~\bibnamefont {Jaccarino}},\ }\href
  {https://doi.org/10.1103/PhysRev.178.804} {\bibfield  {journal} {\bibinfo
  {journal} {Physical Review}\ }\textbf {\bibinfo {volume} {178}},\ \bibinfo
  {pages} {804} (\bibinfo {year} {1969})},\ \bibinfo {note} {publisher:
  American Physical Society}\BibitemShut {NoStop}%
\bibitem [{\citenamefont {Nikotin}\ \emph {et~al.}(1969)\citenamefont
  {Nikotin}, \citenamefont {Lindgård},\ and\ \citenamefont
  {Dietrich}}]{nikotin_magnon_1969}%
  \BibitemOpen
  \bibfield  {author} {\bibinfo {author} {\bibfnamefont {O.}~\bibnamefont
  {Nikotin}}, \bibinfo {author} {\bibfnamefont {P.~A.}\ \bibnamefont
  {Lindgård}},\ and\ \bibinfo {author} {\bibfnamefont {O.~W.}\ \bibnamefont
  {Dietrich}},\ }\href {https://doi.org/10.1088/0022-3719/2/7/309} {\bibfield
  {journal} {\bibinfo  {journal} {Journal of Physics C: Solid State Physics}\
  }\textbf {\bibinfo {volume} {2}},\ \bibinfo {pages} {1168} (\bibinfo {year}
  {1969})}\BibitemShut {NoStop}%
\bibitem [{\citenamefont {Matsubara}\ and\ \citenamefont
  {Inawashiro}(1977)}]{matsubara_mixture_1977}%
  \BibitemOpen
  \bibfield  {author} {\bibinfo {author} {\bibfnamefont {F.}~\bibnamefont
  {Matsubara}}\ and\ \bibinfo {author} {\bibfnamefont {S.}~\bibnamefont
  {Inawashiro}},\ }\href {https://doi.org/10.1143/JPSJ.42.1529} {\bibfield
  {journal} {\bibinfo  {journal} {Journal of the Physical Society of Japan}\
  }\textbf {\bibinfo {volume} {42}},\ \bibinfo {pages} {1529} (\bibinfo {year}
  {1977})},\ \bibinfo {note} {publisher: The Physical Society of
  Japan}\BibitemShut {NoStop}%
\bibitem [{\citenamefont {Bloch}(1966)}]{bloch_103_1966}%
  \BibitemOpen
  \bibfield  {author} {\bibinfo {author} {\bibfnamefont {D.}~\bibnamefont
  {Bloch}},\ }\href {https://doi.org/10.1016/0022-3697(66)90262-9} {\bibfield
  {journal} {\bibinfo  {journal} {Journal of Physics and Chemistry of Solids}\
  }\textbf {\bibinfo {volume} {27}},\ \bibinfo {pages} {881} (\bibinfo {year}
  {1966})}\BibitemShut {NoStop}%
\bibitem [{\citenamefont {Keffer}(1952)}]{keffer_anisotropy_1952}%
  \BibitemOpen
  \bibfield  {author} {\bibinfo {author} {\bibfnamefont {F.}~\bibnamefont
  {Keffer}},\ }\href {https://doi.org/10.1103/PhysRev.87.608} {\bibfield
  {journal} {\bibinfo  {journal} {Physical Review}\ }\textbf {\bibinfo {volume}
  {87}},\ \bibinfo {pages} {608} (\bibinfo {year} {1952})},\ \bibinfo {note}
  {publisher: American Physical Society}\BibitemShut {NoStop}%
\end{thebibliography}%

\end{document}